# Electromagnetic field orientation and dynamics governs advection characteristics within pendent droplets

**Purbarun Dhar** [a,*], **Vivek Jaiswal** [a] and **A R Harikrishnan** [b]

[a] Department of Mechanical Engineering, Indian Institute of Technology Ropar,

Rupnagar–14001, India

[b] Department of Mechanical Engineering, Indian Institute of Technology Madras,

Chennai–600036, India

*Corresponding author*: E–mail: purbarun@iitrpr.ac.in

Phone: +91-1881-24-2173

*Abstract*

The article reports the domineering governing role played by the direction of electric and magnetic fields on the internal advection pattern and strength within salt solution pendant droplets. Literature shows that solutal advection drives circulation cells within salt based droplets. Flow visualization and velocimetry reveals that the direction of the applied field governs the enhancement/reduction in circulation velocity and the directionality of circulation inside the droplet. Further, it is noted that while magnetic fields augment the circulation velocity, the electric field leads to deterioration of the same. The concepts of electro and magnetohydrodynamics are appealed to and a Stokesian stream function based mathematical model to deduce the field mediated velocities has been proposed. The model is found to reveal the roles of and degree of dependence on the governing Hartmann, Stuart, Reynolds and Masuda numbers. The theoretical predictions are observed to be in good agreement with experimental average spatio-temporal velocities. The present findings may have strong implications in microscale electro and/or magnetohydrodynamics.

*Keywords:* Droplet; Particle Image Velocimetry; Electrohydrodynamics; Magnetohydrodynamics; Hartmann number; Masuda number

*Introduction* – Fluid dynamics[1], heat transfer[2, 3] and species transport[4, 5] dynamics in droplets, both pendant[6, 7] as well as sessile[8, 9], has been an area of immense academic importance within the research community. The importance stems from the fact that droplets are essential components in utilities such as thermal management[10, 11], aerospace systems[12, 13], power generation[14], medical devices[15, 16], production engineering[17], etc. In case of multicomponent droplets, such as binary fluids, salt solutions, etc. the thermo-solutal gradient generated across the droplet surface and its bulk is known to drive internal circulation cells[5, 18]. It has been shown in literature that the internal advection behavior is important in modulating fluid dynamics, heat transfer and mass transport from droplets[19, 20] as well as in the ambient air surrounding the drops. The internal advection dynamics within pendant droplets has received wide attention in recent times, since the bulk and interfacial advection dynamics in such droplets is nearly independent of external surfaces (in case of sessile drops). Thereby, internal advection within pendant droplets provides a mimicking picture of flow dynamics within free standing droplets, such as in sprays.

The prospect of electro and magnetohydrodynamic transport in droplets to tune thermofluidic and species transport in microscale flow systems has also received tremendous attention in recent times. It has been shown that presence of electric fields can modify droplet morphology[21], flow dynamics[22], heat transfer[23] as well as mass transport[24] processes. Similar modifications in flow kinetics[25], thermal[26] and species transport[27] phenomena are observed in magnetic droplets under the influence of magnetic field. The presence of the electric and/or magnetic body force leads to polarization force within the fluid, as well as on its interface/ surface, leading to enhanced or deteriorated transport phenomena. Consequently, understanding the electro and magnetohydrodynamic features within droplets is of prime importance, especially for implementations in microscale flow devices and systems. The present article reports the effects of electromagnetic field strength as well as directionality on the nature and strength of internal advection within droplets of polar and/or paramagnetic salt based solutions. It is shown that the advection features is strongly dependent on the field strength as well as direction. A mathematical formulation is proposed to model the experimental observations, and good predictability is observed. The findings may have strong implications in the area of multiphysics of fluid dynamics in microscale flows.

*Experimental methodologies* – The experimental setup used is illustrated in Figure 1 and is very similar as reported by present authors[5, 27]. Sodium iodide (NaI) and Ferric chloride (FeCl$_3$) solutions in water (0.2 M concentration) are employed for polar fluid based electrohydrodynamics and paramagnetic fluid based magnetohydrodynamics, respectively. The selection of salts is based upon previous experiments by the authors[5, 27]. The detailed description of the setup is provided in the supporting information[a].

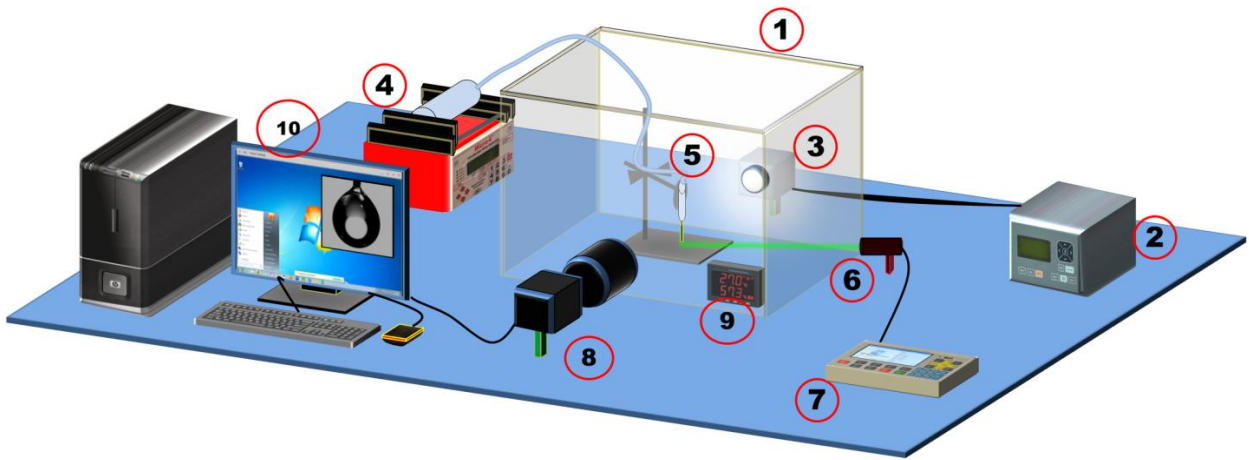

**Figure 1:** Illustration of the experimental setup (1) acrylic enclosure (2) backlight controller (3) LED backlight (4) precision syringe pump (5) dispensing tubing and needle assembly (6) flow visualization laser (7) laser controller (8) CCD camera (9) digitized thermometer and hygrometer (10) data acquisition computer. The pendent drop is positioned between the electric field generator and magnetic field setup as per requirement.

*Results and discussions* – The velocity fields and contours within the in the event of electrohydrodynamics for NaI based droplet have been illustrated in Fig. 2. Fig. 2 (a) shows the velocity field in case of zero-field. Figs. 2 (b) and (c) illustrate the cases for vertically aligned field whereas (d) and (e) illustrate the horizontal field conditions. It is observed that under zero-field condition, the circulation is oriented with the axis of circulation along the plane of the paper (typical direction of flow is shown by the large circular arrow), caused by solute-thermal advection[5]. When the vertical field is applied across the droplet, the flow pattern is observed to change completely in general. At 100 V 50 Hz, the circulation structure is modulated to form two major advection cells, with opposing directionalities, at the bottom hemisphere of the droplet. The field lines in this case towards the top of the pendant, and fan out towards the bulb of the pendant (since the needle is one electrode and the base electrode is a plate electrode placed just below the droplet base). Accordingly, the major region near the bulb of the pendant is within field influence, and hence the bottom of the pendant exhibits major circulation. The force per unit volume (F) on a charged entity due to electromagnetic field is as expressed in Eq. 1 (where $q$, $E$, $\sigma_e$, $v$ and $B$ are the charge, electric field intensity, electrical conductivity of the fluid, bulk velocity of the fluid, and magnetic field strength).

$$\vec{F} = q\vec{E} + (\sigma_e . \vec{E} \times \vec{B}) + \sigma_e (\vec{v} \times \vec{B}) \times \vec{B} \qquad (1)$$

For only electric field condition, the electric body force is directed along the field direction, which redistributes the initial direction of circulation to a vertical form. From the Navier–Stokes equations, the electric body force is known to cause deterioration in the net force on the right hand terms, thereby leading to reduction in the advective left hand components. This essentially causes reduction in the effective circulation velocity.

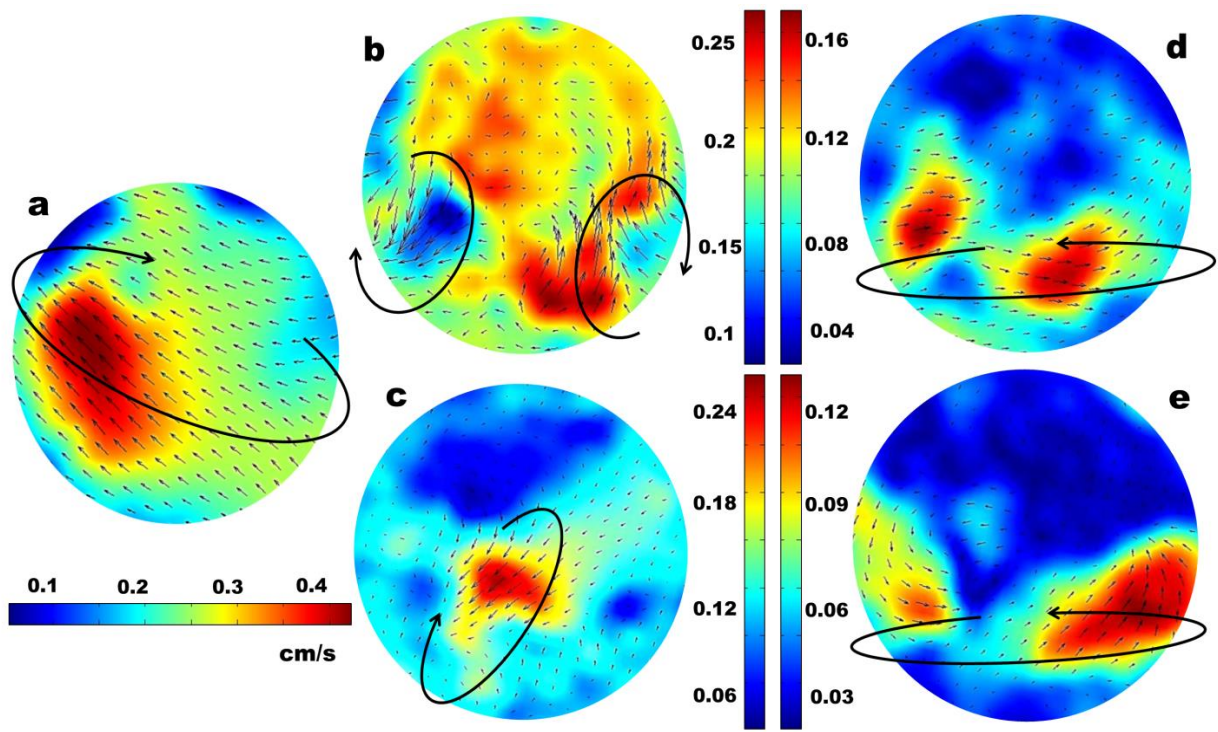

**Figure 2:** Flow dynamics within the pendant drop due to electric field stimulus at a) 0 V b) vertical 100 V 50 Hz c) vertical 100 V 200 Hz d) horizontal 100 V 50 Hz and e) horizontal 250 V 50 Hz. The large arrows illustrate the typical direction of advection. All velocities are in cm/s.

Upon increasing the frequency at the same field strength, it is observed that the circulation velocity is further weakened, resulting in a weak vertical advection cell confined to the bulb of the pendant (fig. 2 (c)). In the case of horizontal field, when 100 V 50 Hz field is applied across the droplet (positioned between two plate electrodes); the direction of circulation is accordingly modulated to the plane passing through the droplet horizontally. This is conjugated with reduction in the average velocity as well (fig. 2 (d)). Interestingly, when the field strength is increased and frequency kept constant (fig. 2 (e)), the reduction in circulation velocity is observed to be less compared to the case when the field strength was kept constant and the frequency was enhanced. This observation also holds true for the vertical field case, and the generic trend observed is that the frequency of the applied field has dominant effect on the strength of internal advection compared to the field strength. However, while the reduction in the right hand side force component of the Navier-Stokes equations can qualitatively explain the reduction in the left hand side advective components, determining the electrohydrodynamic velocities analytically from the Navier-Stokes equations are explicitly difficult. For the theoretical component, the classical electrohydrodynamic formalism for droplets[28] is appealed to. The formalism is based upon the flow within a droplet suspended in an ambient fluid medium (in the present case air) and the flow dynamics within and outside the drop due to an electric field. Since the present external fluid is air (negligibly polar in comparison to water), the external velocity field is scaled to zero and the formalism collapses to only account for the flow within the droplet due to the electric body force. Essentially, the modified body force due to the electric field manifests as a differential pressure gradient across the droplet interface, which in turn drives the internal convective cell.

In a typical spherical system, the Stokes stream function ($r$ and $\theta$ components, with symmetry along $\varphi$ direction) for flow within the droplet is expressible as[28]

$$\psi = \left(C_1 a^{-1} r^3 - C_2 a^{-3} r^5\right) \sin^2 \theta \cos \theta \tag{2}$$

From definition of the Stokes stream function, the $r$ and $\theta$ components of velocity, viz. $u$ and $v$, are determined as

$$u(r,\theta) = \left(C_1 a^{-1} r - C_2 a^{-3} r^3\right)(2\cos^2 \theta - \sin^2 \theta) \tag{3}$$

$$v(r,\theta) = -\left(3C_1 a^{-1} r - 5C_2 a^{-3} r^3\right) \sin \theta \cos \theta \tag{4}$$

In eqns. 2, 3 and 4, the coefficients $C_1$ and $C_2$ are constants which are determined from suitable boundary conditions for the problem. For the scenario at hand, the external flow is negligible, and hence from the balance of the tangential stress components acting at the interface due to interfacial tension and electric force field, the scaled coefficient $C_1$ can be expressed as

$$C_1 = -\frac{9\eta_c}{160\pi f} Md \sin 2\theta \tag{5}$$

Where, $\eta_c$, $f$ and $Md$ are the charge density in the fluid (mass basis), the frequency of the electric field and the Masuda number, respectively. On similar considerations, balance of the normal stress components due to interfacial tension and electric force field yields the relationship between $C_1$ and $C_2$ as

$$3\mu_d(1-3\cos^2\theta)C_1 - \frac{9E_0^2 k_d}{32\pi}\cos^2\theta = C_2 \qquad (6)$$

Where, $\mu_d$, $E_0$ and $k_d$ are the viscosity of the fluid, the electric field strength, and the relative permittivity of the droplet fluid, respectively. The Masuda number (also termed the dielectric electro–Rayleigh number) signifies the ratio of electro-diffusion to momentum diffusion. The generic form of the $Md$ is expressible as[29]

$$Md = \frac{\varepsilon_0 k_d E^2 a^2}{\rho_d v_d^2} \qquad (7)$$

Where, $\varepsilon_0$, $\rho_d$ and $k_d$ are the permittivity of free space, density and kinematic viscosity of the droplet fluid, respectively. The center of the droplet is assumed to be the origin, and the droplet is divided into 180 equi-angular segments. The radial direction is divided into 20 linearly equi-spaced segments, and the intersection of the radial and angular segments gives rise to a grid system (of 3600 grid points) for computations. The eqns. 3-7 are computationally solved simultaneously to determine the two velocity components at each grid point within the droplet. The difference in computational accuracy at higher number of grid points is within ± 3%.

The vector additive velocity at each point is evaluated and the spatially averaged velocity within the whole domain is determined. The spatio-temporally averaged velocity of 600 PIV frames are determined and illustrated in Fig. 3 (a) against the theoretical velocities. It is observed that the present stream function approach is able to accurately predict the experimental velocities within ± 15 %. The dominant role played by the field frequency over the field velocity is also captured by the $Md$ approach in the theoretical model. The sensitivity of the internal advection to field stimulus has been illustrated in fig. 3 (b). In case of water, the advection is found to be weak (~ 0.05 cm/s in general) and field response is naught since there is thermal or solutal advection within the droplet[5, 27]. Since the mean circulation is negligible, the temporal fluctuation in average velocity is also minute. In the event of salt based droplet, thermo-solutal advection leads to strong internal circulation at zero field condition (illustrated in 0–15 s in fig. 3 (b)), with large amplitude temporal variance in average velocity[5]. At 15 s the field is switched on, with the strength and frequency as detailed in the legend in the illustration. As observed from the contours in Fig. 2 and discussed earlier, the typical average velocities are observed to deteriorate. Additionally, the temporal variation in velocity is observed to be reduced drastically, and similar to the case of water, and even lesser in case of high frequencies. As the frequency increases, the ionic population in the fluid experiences faster change in the direction of the field force, which is expected to arrest their electrophoretic and thermal fluctuations to a large extent. Reduction in the electrophoretic drift deteriorates the localized electro-solutal advection velocity within the fluid domain, which leads to reduced fluctuations in the average velocity.

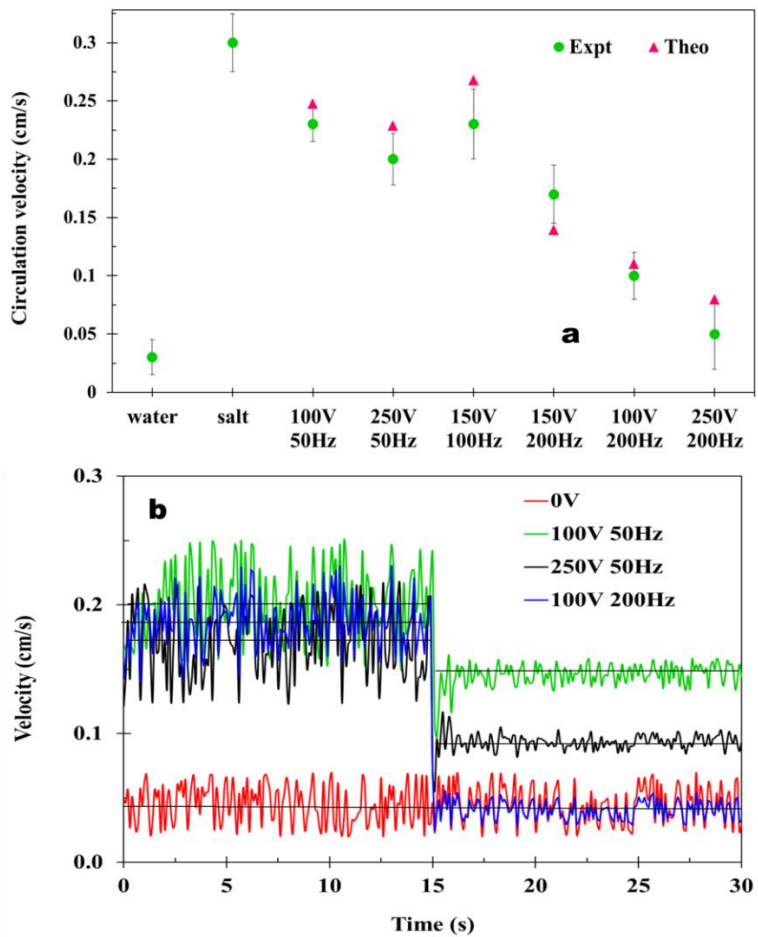

**Figure 3:** (a) Experimental and theoretical circulation velocities (spatio-temporally averaged) for electrohydrodynamic advection within the droplet (b) temporal variation of circulation velocity due to electric field. The straight lines represent the best fit mean velocity. The 0 V red line represents water. For all other cases, salt solution droplet is used. The field is initially at off state and switched on at 15 s.

Fig. 4 illustrates the droplet dynamics in case of magnetic field stimulus. The FeCl$_3$ based paramagnetic salt solution exhibits similar thermo-solutal advection[5] as the NaI based case. In the event of an applied horizontal field, the average velocity of circulation enhances, and the direction of circulation also changes as illustrated in fig. 4 (b) and (c). From eqn. 1, for only magnetic field case, the direction of the body force is evaluated as the vector product of the velocity and field components, which may also be determined from Fleming's rule based on the initial direction of advection at zero-field. Based on the right hand rule, the direction of advection under field stimulus can be determined and is in accordance to fig. 4 (b)–(e). It is noteworthy that the circulation velocity is augmented by the magnetic field, unlike the electric case. The presence of the paramagnetic salt ions induces magnetophoretic drift in the direction of the applied field, which modulates the internal solutal gradient[27], leading to enhanced advection strength. Likewise, in the case of a vertical field, the strength direction of circulation is modulated (fig. 4 (d) and (e)). In the event of vertical 0.24 T field (fig. 4 (d)), the field is not strong enough to induce a single circulation cell within the whole droplet, and the major advection remains confined to the lower bulb of the drop. This advection generates shear within the fluid, leading to a weaker and opposing circulation near the droplet neck. To model the phenomenon, the theory of low Reynolds number magnetohydrodynamic flows is considered[30]. The scaled equations are solved employing balance of shear conditions at the droplet interface and from the zero-field internal circulation velocity. The equation for the subsequent Stokes stream function (for the $r$ and $\theta$ coordinates) is as expressed in Eqn. 8[30].

$$\psi = \sqrt{\frac{C_3 \mu_d}{\rho_d}} r \cos\theta . \frac{1}{\beta} . (1 - e^{-\beta\xi}) \qquad (8)$$

From Eqn. 8, the radial and angular components of velocities at each point can be expressed in terms of Eqns. 9 and 10, respectively.

$$u(r,\theta) = -r^{-1}\sqrt{\frac{C_3 \mu_d}{\rho_d}} + e^{-\beta\xi}\left(r^{-1} + \beta\sqrt{\frac{C_3 \rho_d}{\mu_d}}\cos\theta\cot\theta\right)\sqrt{\frac{C_3 \mu_d}{\rho_d}} \quad (9)$$

$$v(r,\theta) = -r^{-1}\sqrt{\frac{C_3 \mu_d}{\rho_d}}\cot\theta + e^{-\beta\xi}\left(r^{-1}\cot\theta + \beta\sqrt{\frac{C_3 \rho_d}{\mu_d}}\cos\theta\right)\sqrt{\frac{C_3 \mu_d}{\rho_d}} \quad (10)$$

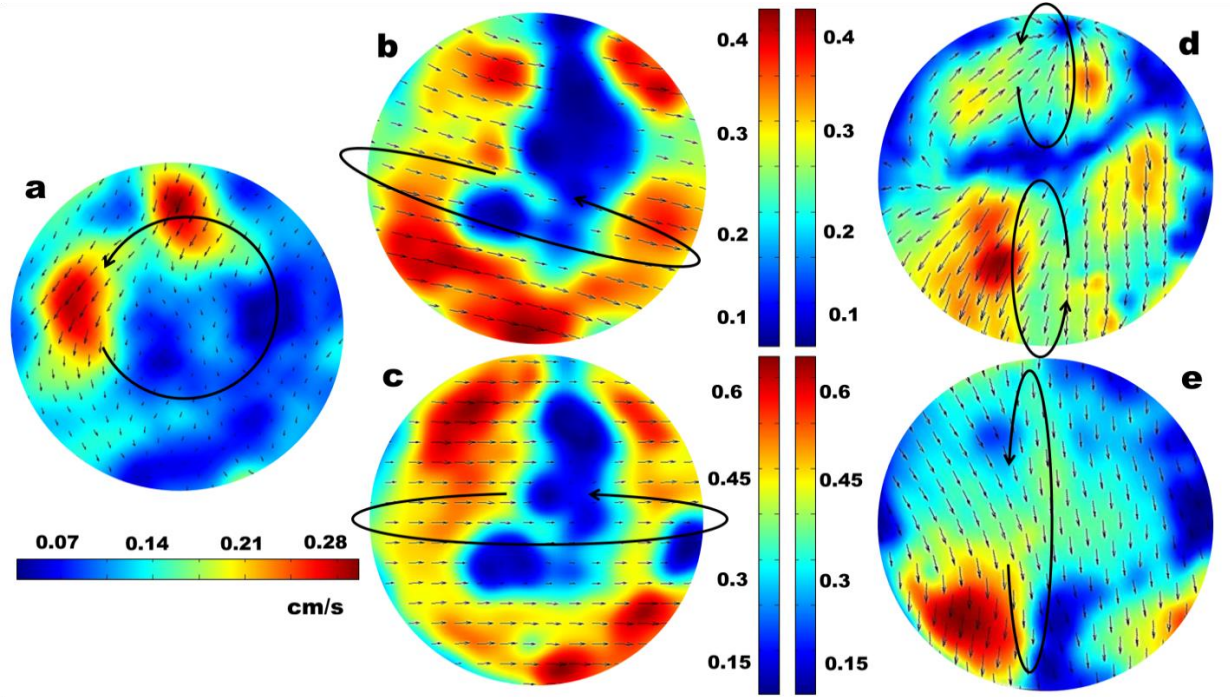

**Figure 4:** Flow dynamics within the pendant drop due to magnetic field stimulus at a) 0 T b) horizontal 0.24 T c) horizontal 0.48 T d) vertical 0.24 T and e) vertical 0.48 T. The large arrows illustrate the typical direction of advection. All velocities are in cm/s.

In eqns. 8–10, the coefficient $C_3$ is a constant which relates the localized velocity components with the coordinates for that location[30]. The variable $\xi$ represents the non-dimensional angular velocity component and is expressible as Eqn. 11. The index $\beta$ represents the non-dimensional form of the magnetohydrodynamic body force acting on the droplet and is expressed as per Eqn. 12.

$$\xi = \sqrt{\frac{C_3 \rho_d}{\mu_d}} r \sin\theta \tag{11}$$

In Eqn. 12, $Re_0$ and $Sr$ represent the internal circulation Reynolds number at zero–field condition and the associated Stuart number, respectively. The latter can be expressed further as in eqn. 13.

$$\beta = \left[1 + \frac{\rho_d v_0^2}{C_3 \mu_d} \frac{Re_0}{Sr}\right]^{\frac{1}{2}} \tag{12}$$

In eqn. 13, $Ha$ represents the Hartmann number governing the magnetic advection with respect to the viscous forces.

$$Sr = \frac{Ha^2}{Re} \tag{13}$$

In the present case, the advection is created by the paramagnetism induced by the $Fe^{3+}$ ions (since non-magnetic salts do not exhibit similar augmented circulations); a modified version of the $Ha$ is employed, viz. $Ha = \rho_d B M_d a / v_0 \mu_d$, where $B$ and $\mathbf{v_0}$ are the magnetic field strength and zero-field circulation velocity. The theoretical velocities have been plotted alongside the experimental velocities (spatially averaged for 600 PIV contours) and accuracy within ± 12 % has been observed. The temporal velocity spectra for the magnetohydrodynamic advection have been illustrated in fig. 5 (b). While the mean velocity is enhanced by magneto-thermosolutal advection[27], there is no such increment or decrease in the temporal fluctuation patterns, unlike the electrohydrodynamic problem. Since the present field is a direct magnetic field, the magnetohydrodynamic force is unidirectional, and thereby has no such effect in arresting or enhancing the local velocity fluctuations.

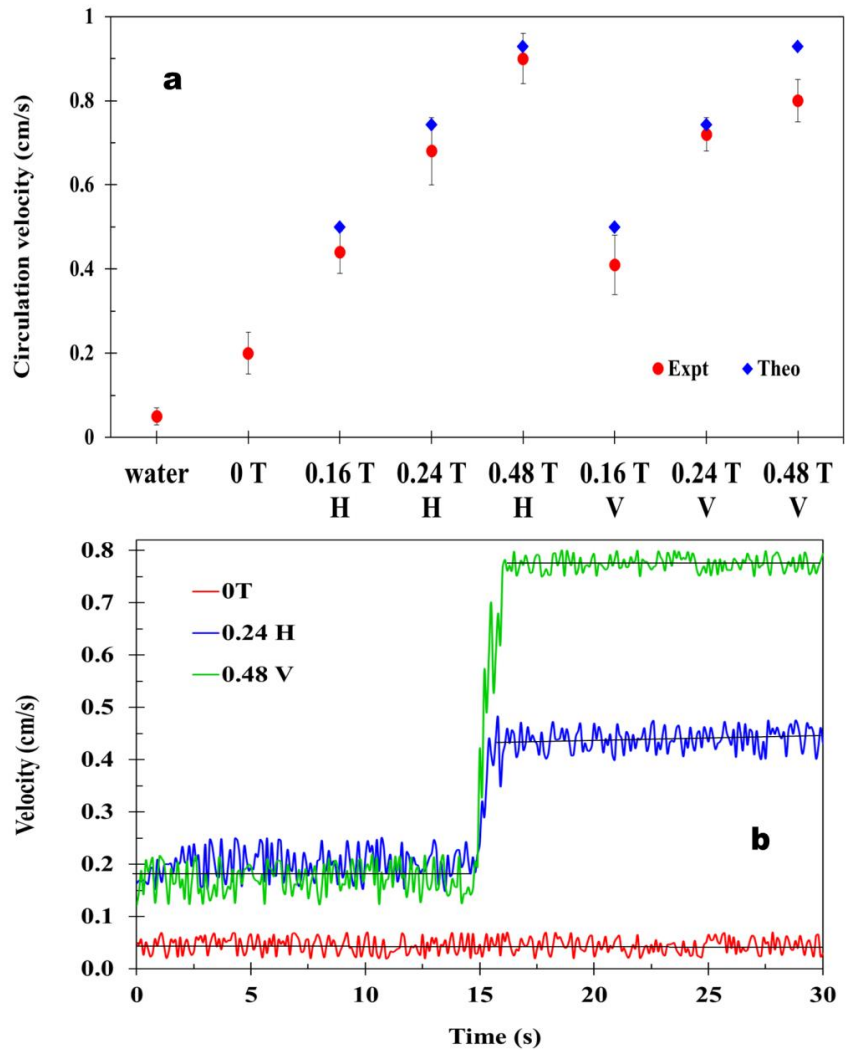

**Figure 5:** (a) Experimental and theoretical circulation velocities (spatio-temporally averaged) for magnetohydrodynamic advection within the droplet (b) temporal variation of circulation velocity due to magnetic field. The straight lines represent the best fit mean velocity. The 0 T red line represents water. For all other cases, paramagnetic salt solution droplet is used. The field is initially at off state and switched on at 15 s.

*Conclusions* – The present article experimentally demonstrates the magnetohydrodynamic and electrohydrodynamic flow behavior within pendant droplets. An experimental setup is employed and PIV is used to visualize flow patterns and quantify flow velocities. For electrohydrodynamics, a salt based droplet is used and horizontal and vertical electrode assemblies are used. It is observed that the field strength, frequency and direction directly influence the flow dynamics. The presence of electric field is observed to arrest the circulation velocity, however, increase in frequency is found to be more dominant in deterioration of velocity than increase in field strength. Dynamic analysis of the velocity spectra with time reveals that the electric field also dampens out temporal oscillations and perturbations in the circulation velocity. The spatially averaged flow velocity has been modelled based on theories of droplet electrodynamics, and accurate predictions are observed. It is shown that the dynamics is strongly influenced by the Masuda number involved. For magnetohydrodynamics, a paramagnetic salt solution is used and the droplet is place in the presence of horizontal or vertical magnetic fields. The magnetic field is observed to augment the velocity of circulations, and also the direction of flow, which can be deduced from vector laws. The magnetic field however is observed to weakly affect the oscillations in the temporal velocity spectrum. The circulation velocity has been modelled based on droplet magnetohydrodynamics, and the predictions are found to be accurate. The dynamics are found to be strongly governed by the Hartmann and Stuart numbers. The present findings may have strong implications towards electromagnetic control and tuning of microfluidic systems.

**Acknowledgement**

PD thanks IIT Ropar for the financial support towards the present research (grants IITRPR/Interdisciplinary/CDT and ISIRD IITRPR/Research/193)